# An intercomparison of monthly surface air temperature on islands and proximate moorings across the tropical Indo–Pacific

Kristopher B. Karnauskas [1,2,3,4,*], Jeffrey P. Donnelly [4], and Kevin J. Anchukaitis [5]

[1] Cooperative Institute for Research in Environmental Sciences, University of Colorado Boulder
[2] Department of Atmospheric and Oceanic Sciences, University of Colorado Boulder
[3] Department of Environmental and Occupational Health, Colorado School of Public Health
[4] Department of Geology and Geophysics, Woods Hole Oceanographic Institution
[5] School of Geography and Development, University of Arizona
* Corresponding author (kristopher.karnauskas@colorado.edu)

**Abstract**

Despite the importance and vulnerability of small island nations and their ecosystems, they frequently have insufficient observations to provide baseline or long–term perspectives on climate variability and change, and global model experiments rarely have the resolution to include them. Many studies in observational climatology, climate modeling, and paleoclimate thus depend to varying degrees on an approximation equating near–surface marine and terrestrial island climates, often with direct implications for island societies and resources. Here we investigate the validity of this approximation, and by extension the viability of offshore moorings to serve as proxies for island climate variability, by comparing monthly mean observations on a diverse set of 17 islands across the global tropics with those from ocean moorings proximate to each island. While some island–mooring pairs exhibit a mean offset in surface air temperature, these cannot be explained as a simple function of one parameter including distance from the mooring, station elevation, or island dimensions. Overall, the seasonal to interannual variability in near–surface climate at monthly time scales at meteorological stations on tropical islands is captured remarkably well by moorings positioned up to 1,000 km offshore.

## 1. Introduction

There are substantial challenges to characterizing observed variability and trends in surface air temperature on tropical islands, despite potentially significant ecological and societal impacts (Giambelluca et al. 2008). Chief among them concerns the length and consistency of meteorological observations on such islands, if they exist at all. Possible alternatives to serve as proxies for island air temperature include sea surface temperature (SST) measurements (Stephenson et al. 2008) and marine air temperature measurements made by ships (Folland et al. 2003). The use of mooring observations has yet to be explored in this context. There would be several potential advantages if climate records from moorings could be utilized in this way, including their stationarity (in space) and, in some cases, their length and completeness.

The validity of using marine near–surface climate to make inferences about island climates is also important for studies using global climate models (GCMs) to predict future terrestrial climate, hydrology, and ecosystem changes. In other words, can the near–surface climate on a sub–grid scale tropical island (*i.e.*, one that does not exist in GCMs, given their resolution) be approximated by a GCM grid cell where there is only open ocean? This approximation is made implicitly in multi–model assessments and projections such as the periodic Assessment Reports of the Intergovernmental Panel on Climate Change (IPCC) (Solomon et al. 2007; Taylor et al. 2012). Pairs of moorings and islands that are close to one another provide a unique opportunity to evaluate this approximation because the moorings are, in effect, observing what GCMs are simulating, while measurements made on the islands represent the truth.

Finally, the extent to which distal marine environments reflect nearby islands also has implications for interpreting some marine–based paleoclimate archives as proxies for island climate (Field and Lape 2010), and for interpreting terrestrial or lacustrine paleoclimate archives developed on islands as proxies for the broader marine environment (Conroy et al. 2009; Jacoby et al. 2004). For all of these reasons, this study explores the viability of offshore moorings to serve as proxy for island climate variability. The sources of observations and their treatment in this study are described in section 2, the main results are presented in section 3, and a discussion is given in section 4.

## 2. Data and Methods

Monthly mean surface air temperature observations from island stations were gathered from the U.S. National Oceanic and Atmospheric Administration (NOAA) National Climatic Data Center (NCDC) Global Historical Climatology Network version 3 (GHCN v3) (Lawrimore et al. 2011). Moored surface air temperature measurements were obtained from the Tropical Atmosphere–Ocean (TAO) array (McPhaden et al. 1998) coordinated by the NOAA Pacific Marine Environmental Laboratory (PMEL) and the Japan Agency for Marine–Earth Science and Technology (JAMSTEC), the Research Moored Array for African–Asian–Australian Monsoon Analysis and Prediction (RAMA) array (McPhaden et al. 2009), and the Woods Hole Oceanographic Institution (WHOI)–University of Hawaii Ocean Time Series (WHOTS) mooring. Air temperature data from TAO and RAMA moorings were provided directly as monthly means from the NOAA/PMEL data portal, and monthly means were computed from the high–frequency (1 minute) measurements by the WHOTS mooring as provided by the WHOTS project data portal.

The island stations and moorings used in this study are listed along with several details in Table 1 and mapped in Fig. 1. The 17 island stations (14 in the tropical Pacific Ocean including four in the U.S. state of Hawaii, and three in the tropical Indian Ocean) were selected for inclusion based on having at least two years of overlapping measurements of monthly mean surface air temperature from a meteorological station on the island and from at least one mooring positioned within 1,000 km of the island station. Six islands are in similar proximity to two moorings; in those cases, monthly mean data from the two moorings were averaged to create a single moored time series. Given the distribution and scientific focus of the TAO array—monitoring the El Nino–Southern Oscillation in the equatorial Pacific (McPhaden et al. 1998)—the Pacific sites are primarily within ~10° latitude of the equator, in addition to Hawaii near the northern edge of the tropics. The equatorial Pacific mooring data extend back to as early as 1985 (aligned with the initial deployment of the TAO array), while the Hawaiian data begins in 2004 (when the WHOTS moorings became operational). The Indian Ocean data sets begin in the late 2000s, with the deployment of the RAMA array, and are mostly complete through to present (early 2015). The closest proximity for an island–mooring pair is the island of Kiritimati and the mooring that was operating within 50 km thereof for two brief periods of time in the late 1980s. Funafuti Atoll in Tuvalu is also relatively close to a mooring (100 km). The most remote island–mooring pair included is Tarawa Atoll (in the Gilbert Islands, Republic of Kiribati) and the two near–equatorial moorings roughly 890 km to the west at 165°E. The majority of the remaining island–mooring pairs are separated by a few hundred kilometers. Unfortunately, no islands from the Atlantic Ocean can be included as there are none with surface air temperature observations coincident and proximate to Prediction and Research Moored Array in the Atlantic (PIRATA) moorings (Bourles et al. 2008).

Given the diversity of landscapes and geologic history of the 17 islands included in this study, it is useful to try to generalize them as types that are meaningful for the present analysis. For example, one might hypothesize that a mooring better represents the near–surface climate of a low–lying coral atoll (*e.g.*, Kwajalein, Republic of Marshall Islands) than a large, mountainous (volcanic) island such as Kauai, Hawaii (U.S.). The bivariate distribution of the horizontal and vertical scales of the islands included in this study, represented by the major axis and maximum elevation of each island, is shown in Fig. 2. The four true atolls included in this study (Kwajalein, Tarawa, Tuvalu, and Cocos) are very long (~100 km), narrow (~1 km), and low–lying (a few meters above sea level) strips of land, and are therefore not shown on Fig. 2 and simply called type 1. Otherwise, Fig. 2 illuminates a natural separation between type 2 islands that are both small in horizontal scale (<100 km along the major axis) and low lying (<500 m maximum elevation), and larger islands (type 3). Although Kiritimati (a.k.a. Christmas Island, Republic of Kiribati) is also an atoll in the geologic sense, it features an exceptionally broad emergent land area relative to its lagoons and is therefore designated type 2. The outlier in Fig. 2 (250 km long but only 457 m tall) is Andaman in the Bay of Bengal. The three tallest islands are Big Island/Island of Hawaii, Maui, and Kauai; Oahu and Hiva Oa, Marquesas tie for the fourth tallest.

Three metrics are used for intercomparison of monthly surface air temperature between pairs of island stations and moorings: mean bias, temporal correlation, and the ratio of standard deviations over the full overlapping periods. The mean bias is computed as the mean difference between the mooring and the island station (mean[$T_{mooring}-T_{island}$]). The temporal correlation is computed as the linear Pearson's product moment correlation coefficient, and the standard deviation ratio is the ratio of the temporal standard deviation of the mooring time series to that of the island station time series ($\sigma_{mooring}/\sigma_{island}$). For appropriate sites, the mean climatological seasonal cycles of surface air temperature are also compared including their amplitudes and phasing.

## 3. Results

Temporal correlations between pairs of moorings and island stations across the tropical Pacific (not including Hawaii) and Indian Ocean (Figs. 3, 5e) range from 0.17 (Banaba) to 0.80 (Kiritimati and Seychelles), and mean biases range from −1.1°C (Kiritimati) to 0.0 (Tuvalu) to 1.1°C (Marquesas) (Figs. 3, 5b). The correlations between all time series pairs in Figs. 3 (and summarized in Fig. 5b), except for Banaba and Marquesas, are statistically significant at the 95% confidence level, accounting for the effective number of degrees of freedom (Bretherton et al. 1999). Seven pairs exhibit both a high correlation ($r>0.5$) and a small bias (|bias|<0.5°C), including the Pacific islands of Palau, Kwajalein, Tarawa, Tuvalu, and all three islands in the

Indian Ocean (Seychelles, Andaman, and Cocos). Interestingly, the pair with the lowest correlation (Banaba) also has a small mean bias (0.13°C) despite 510 km distance between the mooring and island station (see discussion below). Marquesas has both a large bias (1.1°C) and a relatively low temporal correlation ($r$=0.29). The standard deviations are quite similar between the islands stations and moorings (Fig. 3, 5h)—generally between 0.8 and 1.1. Outliers include Chuuk with larger variability on the island (0.63) and Seychelles with larger variability at the mooring (1.4), the latter of which is clearly explained by differences in the amplitude of the seasonal cycle.

Temporal correlations between the Hawaiian stations and the WHOTS mooring are all higher than the highest among the other 13 sites found deeper within the tropics—between 0.85 and 0.90 (Fig. 4, 5f), while the mean biases fill the range of the other 13 sites (Fig. 5c). The largest mean bias is between Oahu and the WHOTS mooring (–1.4°C), likely related to Honolulu being on the warmer, leeward side of Oahu, and the station being located on heavily developed land at a major international airport. Likewise, the opposite bias between Big Island/Island of Hawaii and the mooring is likely a function of the station location on the cooler, windward side. The absolute mean biases on Kauai and Maui are very small (~0.2°C). As the high correlations between the Hawaiian stations and the WHOTS mooring mentioned above are clearly benefitting from the high amplitude, nearly synchronous annual cycles, time series of the deseasonalized surface air temperature anomalies (mean monthly climatology removed) at the WHOTS mooring, Kauai, and Oahu are also shown in Fig. 4. Temporal correlations between the anomaly time series are 0.55 and 0.64 for Kauai and Oahu, respectively (Fig. 5f), both are significant at the 95% confidence level, and there is negligible change in the standard deviation ratio of the anomalies relative to the full time series (Fig. 5i).

In order to compare the climatological mean seasonal cycles of surface air temperature between islands and moorings, three criteria must be met: (*1*) a robust seasonal cycle must be inherently present in the region (*i.e.*, distinguishable from interannual variability), (*2*) long and complete enough observational records must exist at both the island and mooring to reliably establish the mean seasonal cycles, and (*3*) there must be only a minor distance in latitude between the island and mooring. Seven island–mooring pairs meet these criteria: Palau and Yap in the deep tropical Pacific, Andaman and Cocos in the tropical Indian Ocean, as well as Kauai, Oahu, and Big Island/Island of Hawaii (Figs. 7–8). For Palau and Yap, both near 8°N, 137°E, the climatological mean seasonal cycles are correlated with the mooring climatology >0.8, the maximum correlation is at zero lag, and in both cases the mooring has a slightly larger amplitude (~15%). At Andaman and Cocos in the tropical Indian Ocean, the correlations with the mooring climatologies are 0.64 and 0.94, respectively, and in both cases this is the maximum lag found with the island leading by one month; amplitudes between the islands and moorings differ by 17%–38%. At the Hawaiian islands of Kauai, Oahu, and Big Island/Island of Hawaii (Fig. 8), the correlations with the mooring climatology are between 0.96–0.99, with the islands leading the mooring by one month. The one–month lag of the climatological annual cycles at the moorings relative to the islands may be explained by the marine atmospheric boundary layer being in quasi–equilibrium with the ocean surface at monthly time scales, and the ocean mixed layer having a slower response to insolation. It is also interesting that a robust lead–lag relationship does not appear at the western Pacific sites of Palau and Yap, which may be related to feedbacks between warm pool convection and radiation.

## 4. Discussion

Assuming perfect observations, mean biases could be related to any combination of three factors: (*1*) the island and mooring being displaced across a mean horizontal gradient of surface air temperature, (*2*) the island station being positioned at high enough elevation to record systematically cooler temperatures than the mooring, and (*3*) the island station being positioned within a microclimate that is significantly influenced by the island itself, including orographic influence on circulation, modified surface downwelling shortwave radiation due to cloud cover, and/or modified surface turbulent fluxes and longwave emission (a literal "heat island" effect—normally applied to urban areas). For example, an island may generate a local warm anomaly, but be displaced across horizontal temperature gradient toward a cooler mean climate. This may explain the small mean bias at Banaba (0.13°C), as that island is a relatively flat, homogeneous surface situated ~5° longitude east of the moorings with which it is paired, and therefore toward cooler climatological temperatures. Alternatively, an island station may be at altitude but the island displaced across a horizontal gradient toward a warmer mean climate; all of these situations would lead to an effective cancelation and apparent reduction of mean bias.

While it is not feasible to control for each of these parameters independently with only 17 diverse sites distributed across several distinct climate regimes, the strongest control on mean bias and temporal correlation is station elevation (Fig. 5a, d). For all of the island stations at elevation greater than 20 m above sea level (not including Hawaiian sites), the mean surface air temperatures are indeed systematically cooler than at the mooring, while biases may be of either sign below 20 m. At the Hawaiian sites, whether the station is on the leeward or windward side appears to overwhelm the influence of station elevation. Island type appears to influence mean bias (Fig. 6a); the smallest absolute mean biases are associated with atolls, and incrementally larger absolute mean biases are associated with type 2 and 3 islands. However, the sign of the mean bias may be of either sign, likely depending on a combination of the aforementioned factors. Surprisingly, when considering all 17 islands including Hawaiian, the island type does not determine the strength of the temporal correlation between island stations and mooring temperatures (Fig. 6b); the correlations averaged across type 3 islands are comparable to those for atolls.

Overall, the near–surface climate at monthly time scales at meteorological stations on tropical islands is captured remarkably well by moorings positioned up to 1,000 km offshore. Surprisingly, this includes the relatively very large Hawaiian Islands. Except on atolls, which have negligible offset relative to the moorings, the mean biases for larger islands are not a simple function of any single parameter such as station elevation, position relative to topography, island dimensions, or distance from the mooring. Depending on whether it is the mean absolute climate or the variability/change that is of interest, a case–by–case investigation may be necessary to establish the suitability of a moored record as a proxy for island climate, the applicability of a GCM simulation to a particular island climate, an island–based proxy record to the broader marine climate, or a nearby marine–based proxy record to the island climate.

## Acknowledgements

The authors thank NOAA NCDC for providing GHCN station observations and NOAA PMEL for providing TAO and RAMA mooring observations. This publication is also based upon observations from the WHOI–Hawaii Ocean Timeseries Site (WHOTS) mooring, which is supported by NOAA through the Cooperative Institute for Climate and Ocean Research (CICOR) under Grant No. NA17RJ1223 and NA090AR4320129 to WHOI, and by NSF grants OCE–0327513, OCE–752606, and OCE–0926766 to the University of Hawaii. K.B.K. and J.P.D. acknowledge support from the Strategic Environmental Research and Development Program (SERDP). K.B.K. further acknowledges support from the Alfred P. Sloan Foundation and the James E. and Barbara V. Moltz Fellowship administered by the WHOI Ocean and Climate Change Institute (OCCI). K.J.A. acknowledges support from NSF grant BCS–1263609. The authors thank Al Plueddemann for assistance and insightful discussions pertaining to the moored WHOTS observations. All data analyzed in this paper are publicly available online. GHCN v3 station observations are available via FTP at ftp.ncdc.noaa.gov/pub/data/ghcn/v3/, TAO and RAMA mooring observations are available for download at http://www.pmel.noaa.gov/tao/disdel/disdel.html, and WHOTS mooring observations are available for download at http://uop.whoi.edu/projects/WHOTS/whotsarchive.html.

| Island* | Coords. | Elev. | Type | Moorings (Array) | Dist. | Overlap | $N_{mon.}$ |
|---|---|---|---|---|---|---|---|
| Palau | 7.3, 134.5 | 33 | 2 | 5, 137 \| 8, 137 (T) | 290 | 1993–2014 | 141 |
| Yap | 9.5, 138.1 | 17 | 2 | 8, 137 (T) | 200 | 1995–2013 | 118 |
| Chuuk | 7.5, 151.9 | 2 | 2 | 5, 156 \| 8, 156 (T) | 460 | 1991–2014 | 241 |
| Pohnpei | 7.0, 158.2 | 46 | 3 | 5, 156 \| 8, 156 (T) | 270 | 1991–2014 | 237 |
| Kwajalein | 8.7, 167.7 | 8 | 1 | 8, 165 (T) | 310 | 1989–2014 | 244 |
| Banaba | –0.9, 169.5 | 66 | 2 | –2, 165 \| 0, 165 (T) | 510 | 1985–1990 | 55 |
| Tarawa | 1.4, 172.9 | 4 | 1 | 0, 165 \| 2, 165 (T) | 890 | 1985–1998 | 115 |
| Tuvalu | –8.5, 179.2 | 2 | 1 | –8, 180 (T) | 100 | 1994–2010 | 104 |
| Kiritimati | 2.0, –157.5 | 3 | 2 | 2, –157 (T) | 50 | 1985–1990 | 31 |
| Marquesas | –9.8, –139.0 | 52 | 3 | –5, –140 (T) | 540 | 1990–2014 | 280 |
| Seychelles | –4.7, 55.5 | 3 | 3 | –8, 55 (R) | 370 | 2008–2013 | 57 |
| Andaman | 11.7, 92.7 | 79 | 3 | 12, 90 (R) | 300 | 2007–2014 | 66 |
| Cocos | –12.2, 96.8 | 3 | 1 | –8, 95 \| –12, 93 (R) | 420 | 2009–2014 | 59 |
| Kauai | 22.0, –159.4 | 45 | 3 | 22.75, –158 (W) | 160 | 2004–2014 | 122 |
| Oahu | 21.4, –157.9 | 5 | 3 | 22.75, –158 (W) | 160 | 2004–2014 | 123 |
| Maui | 20.9, –156.4 | 20 | 3 | 22.75, –158 (W) | 260 | 2004–2014 | 41 |
| Big Island | 19.7, –155.1 | 11 | 3 | 22.75, –158 (W) | 450 | 2004–2014 | 91 |

**Table 1.** Complete list of island stations included in this study by name (*see geographical notes below), latitude (°N) and longitude (°E), station elevation (m), and type of island. Atolls are type 1, islands with major axis <100 km and max. elevation <500 m are type 2, and larger islands are type 3. Also listed are the proximate moorings used for intercomparison, the distance between each island station and the nearest mooring (km), the full temporal range of overlapping data, and the number of months ($N_{mon}$) within that range with data. Array "T" stands for the NOAA/PMEL Tropical Atmosphere Ocean (TAO) array of moorings in the tropical Pacific Ocean, "R" for the Research Moored Array for African–Asian–Australian Monsoon Analysis and Prediction (RAMA) array of moorings in the tropical Indian Ocean, and "W" for the WHOI/University of Hawaii Ocean Time Series (WHOTS) mooring north of Hawaii.

* *Palau* (Koror Island) is part of the Republic of Palau; *Yap*, *Chuuk* (Island of Weno inside Chuuk Lagoon of Chuuk Atoll), and *Pohnpei* are part of the Federated States of Micronesia; *Kwajalein* Island on Kwajalein Atoll is part of the Republic of the Marshall Islands; *Banaba* Island, *Tarawa* Atoll (Gilbert Islands), and *Kiritimati* (Line Islands) are part of the Republic of Kiribati; *Tuvalu* (Funafuti Atoll) is part of Tuvalu; *Marquesas* (City of Atuona on Hiva Oa Island) is part of French Polynesia; *Seychelles* (City of Victoria on Mahé Island) is part of the Republic of Seychelles; *Andaman* (City of Port Blair on South Andaman Island) is part of the Andaman and Nicobar Islands, India; *Cocos* (West Island) is part of the Territory of the Cocos (Keeling) Islands, Australia. The Hawaiian (U.S.) stations are Lihue on *Kauai*, Honolulu on *Oahu*, Kahului on *Maui*, and Hilo on Hawaii or *Big Island*.

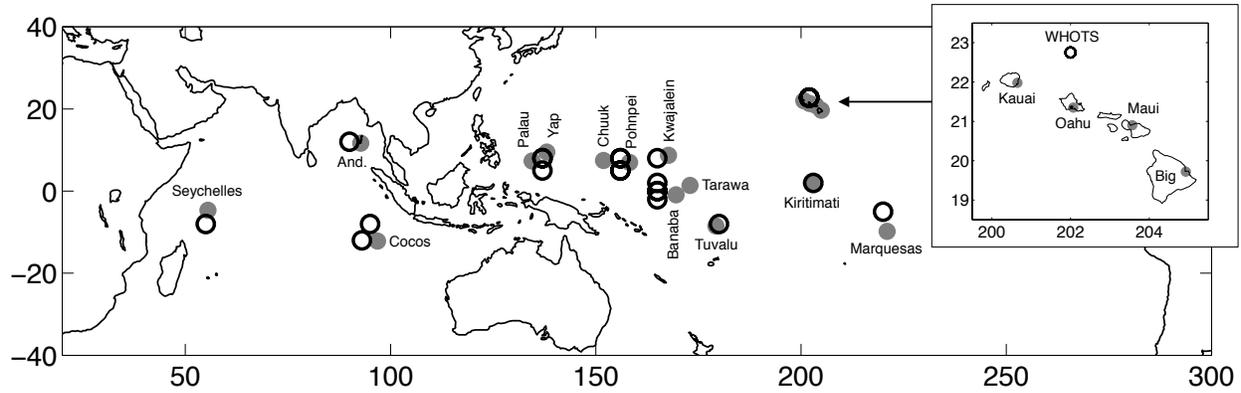

**Figure 1.** Overview map of island stations and proximate moorings included in this study. Gray filled circles indicate islands and open black circles indicate moorings (RAMA array in the tropical Indian Ocean, TAO array in the tropical Pacific Ocean, and the WHOTS mooring north of Hawaii). The inset provides more detail on the four island stations located in the U.S. state of Hawaii and the WHOTS mooring.

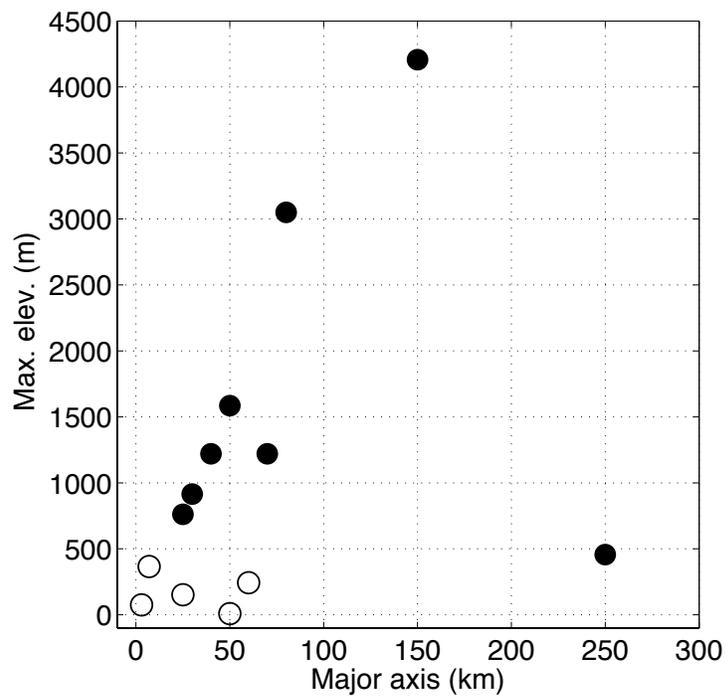

**Figure 2.** Scatterplot of major axis (km) versus max. elevation (m) for the islands included in this study (not including atolls/type 1). Type 2 (3) islands are indicated by open (filled) circles. The criteria for type 2 is major axis <100 km and max. elevation <500 m.

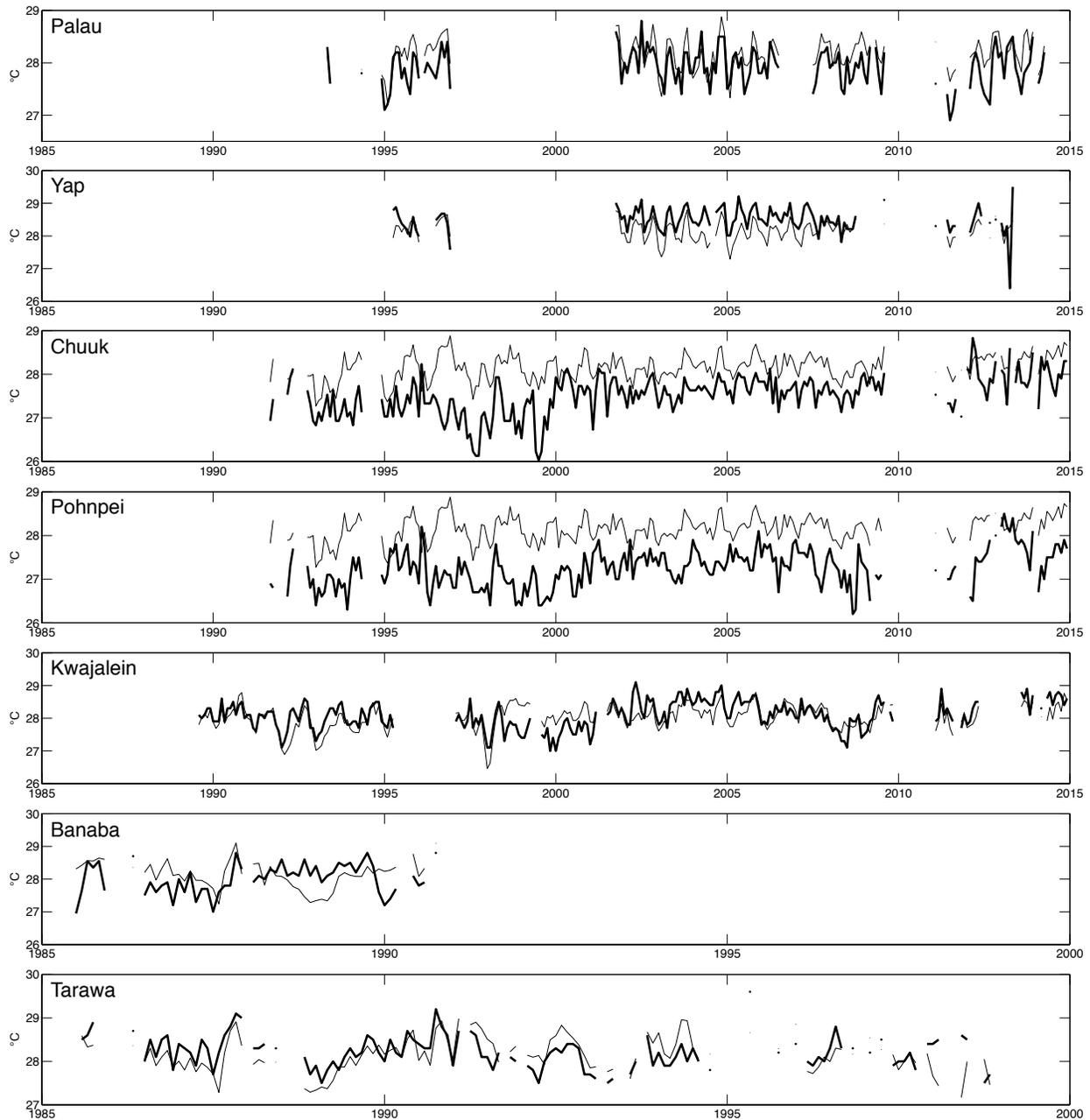

**Figure 3.** Time series of monthly mean surface air temperature (°C) at the 13 tropical Indo–Pacific islands not including Hawaii (thick lines) and proximate TAO or RAMA moorings (thin lines). Refer to Table 1 and Fig. 1 for more information on each site, and Fig. 5 for a summary of statistics.

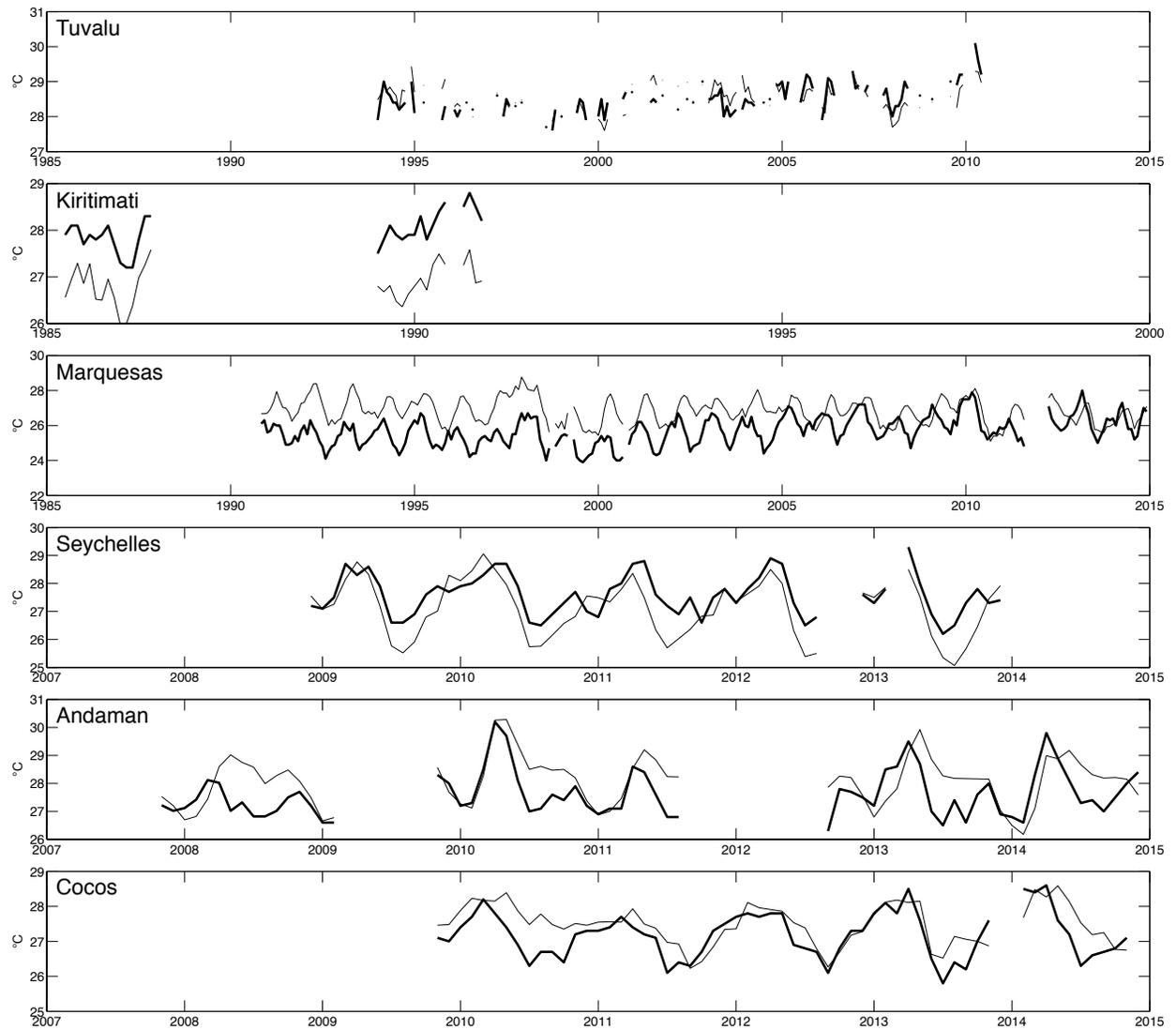

**Figure 3 (continued).**

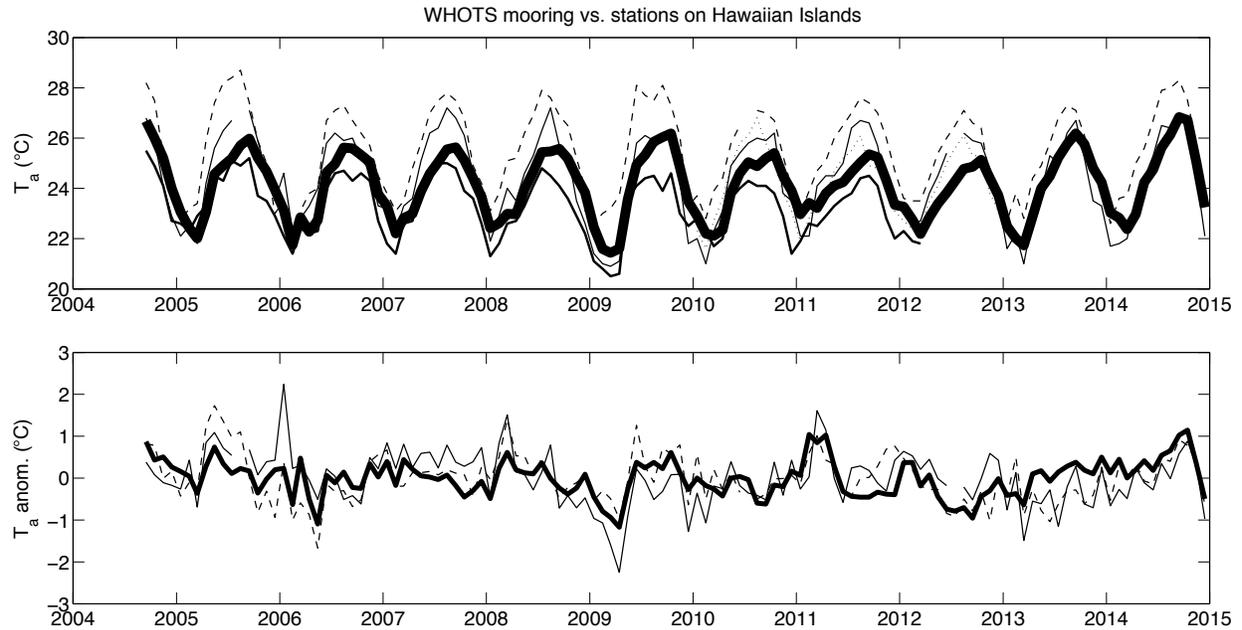

**Figure 4.** Top: Time series of monthly mean surface air temperature (°C) at the WHOTS mooring (thickest solid) and the U.S. Hawaiian island stations on Kauai (thinnest solid), Oahu (dashed), Maui (dotted), and Big Island/Island of Hawaii (medium thickness). Correlation coefficients are 0.89, 0.87, 0.85, and 0.90 for the WHOTS mooring versus Kauai, Oahu, Maui, and Big Island/Island of Hawaii, respectively. Bottom: As in top, but for anomalies (*i.e.*, mean monthly climatology removed). Only Kauai (thinnest solid) and Oahu (dashed) are included as those stations have the longest complete records, extending through 2014. Correlation coefficients are 0.55 and 0.64 for the WHOTS mooring versus Kauai and Oahu, respectively.

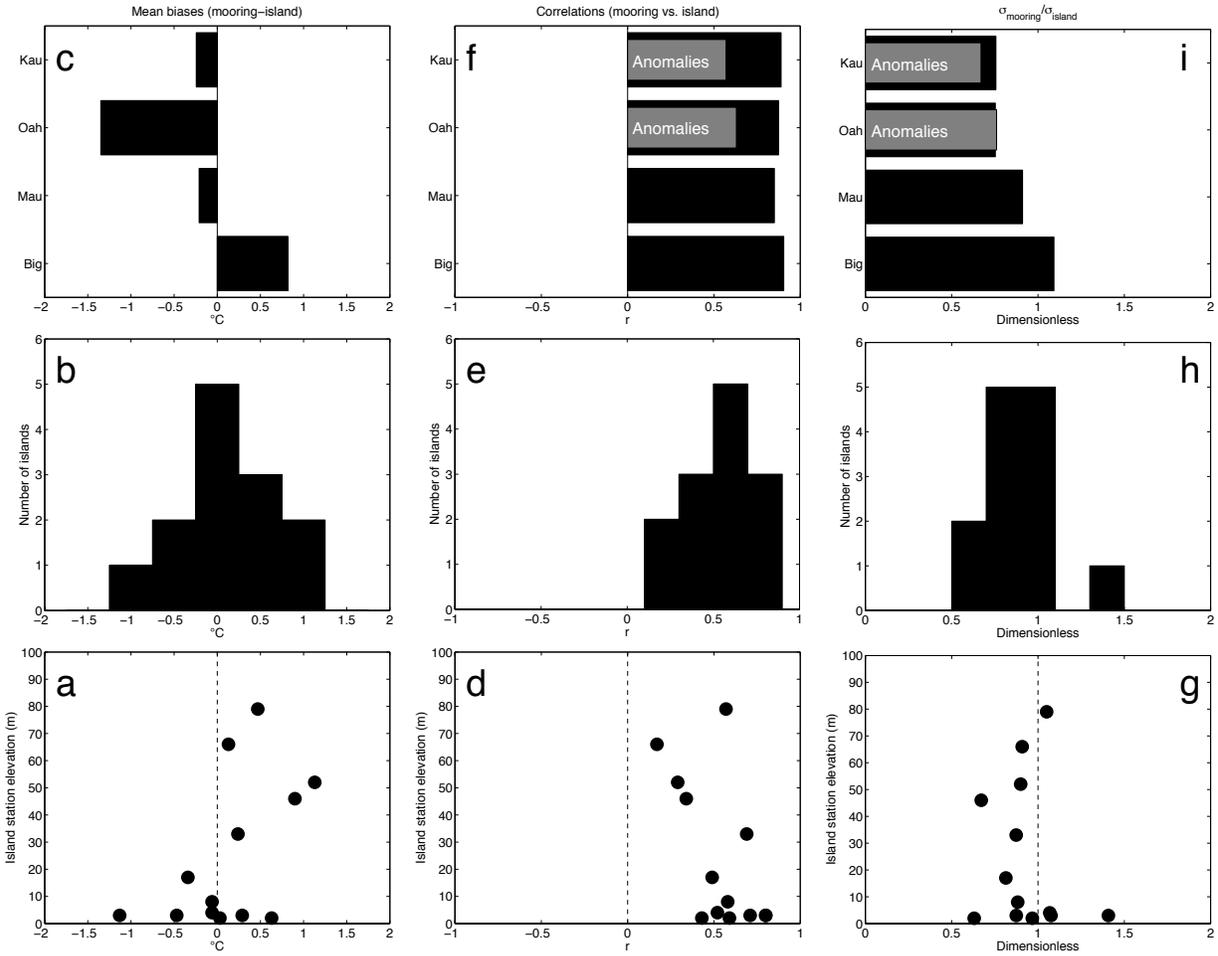

**Figure 5.** Summary of intercomparison of monthly mean surface air temperature at island stations and proximate moorings. Mean bias (mooring–island; °C) as a function of island station elevation for the 13 tropical Indo–Pacific sites not including Hawaii (a), histogram of mean biases for the 13 tropical Indo–Pacific sites not including Hawaii (b), and mean biases for the four Hawaiian island stations (c). (d–f) As in (a–c) but for the correlation coefficient between island stations and moorings. (g–i) As in (a–c) but for the ratio of standard deviations ($\sigma_{mooring}/\sigma_{island}$; dimensionless). In panels f and i, the gray bars indicate results for the anomalies (*i.e.*, mean monthly climatology removed from both the island station and mooring time series).

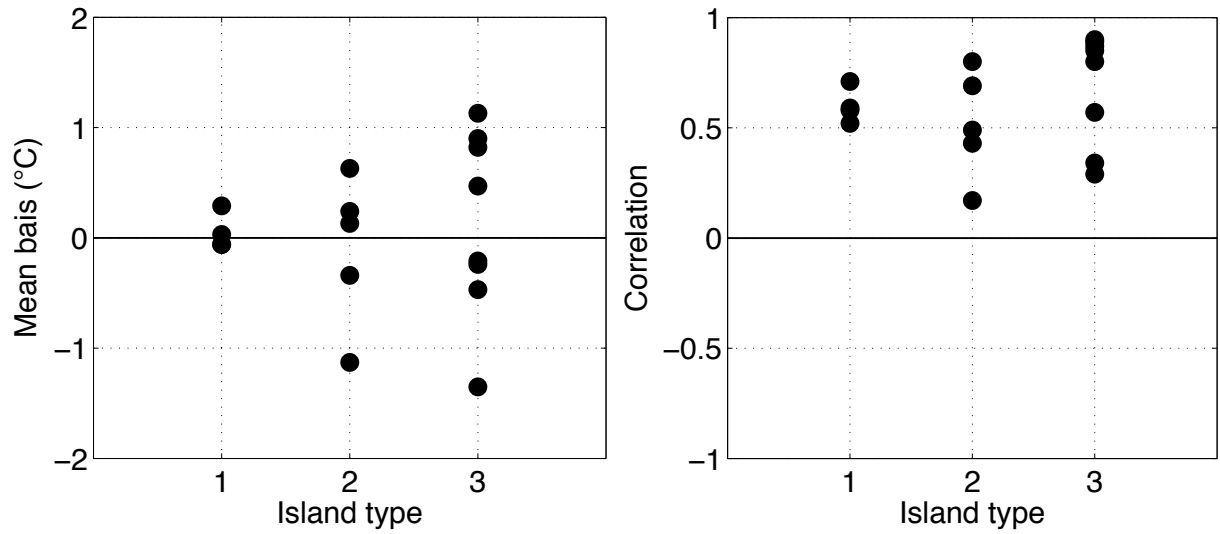

**Figure 6.** Scatterplots of mean bias (°C) and correlation as a function of island type for all 17 island–mooring pairs in this study including the four Hawaiian sites. Refer to Table 1 and section 2 of the main text for an explanation of island type.

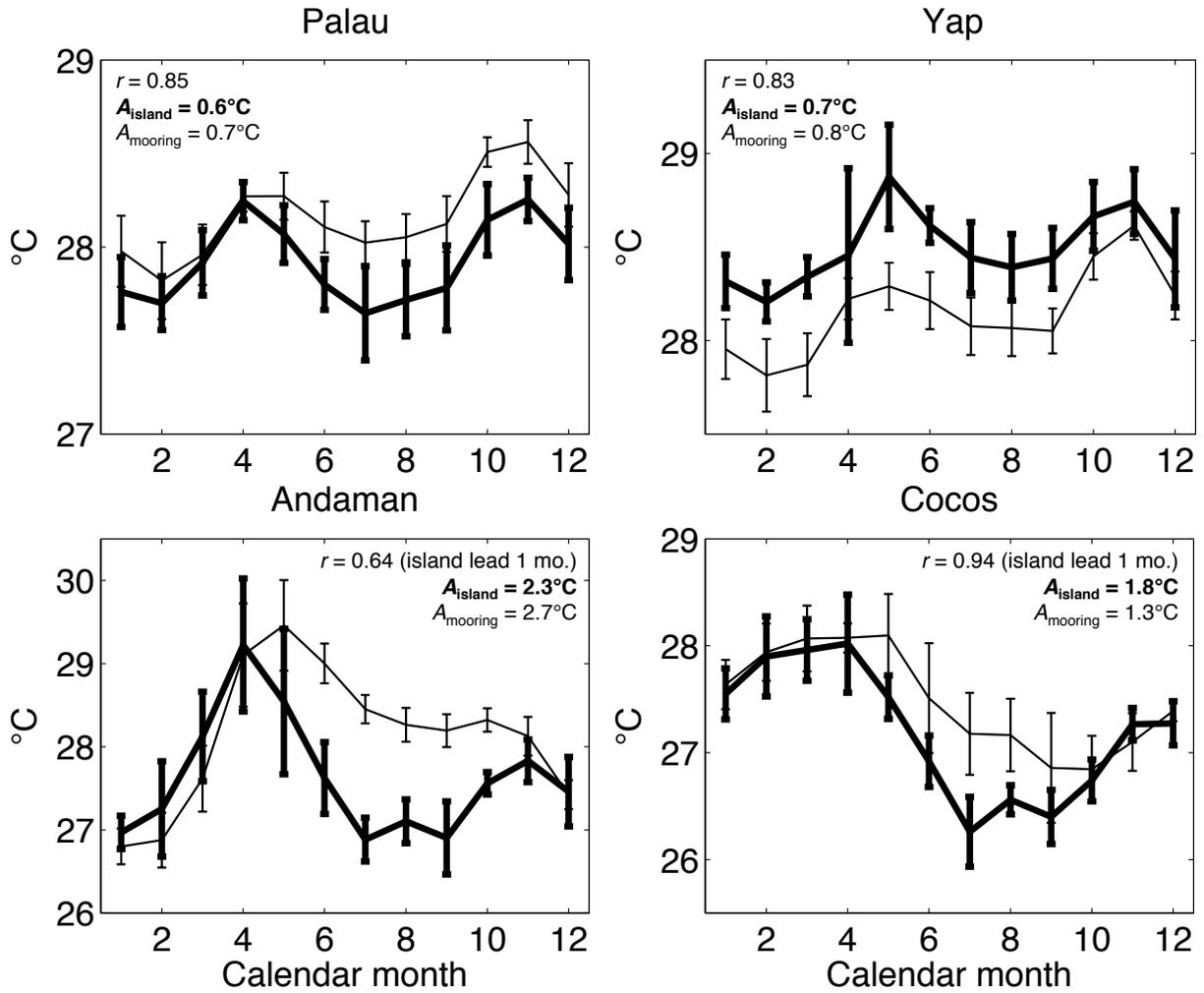

**Figure 7.** Climatological mean seasonal cycles of monthly surface air temperature (°C) at the islands of Palau, Yap, Andaman, and Cocos (thick lines), and the moorings with which they are paired (thin lines; see Table 1). Error bars represent +/– 2 standard errors of the mean. Noted in the corner of each panel is the correlation coefficient between the island and mooring seasonal cycles (including if the maximum correlation is found at a lagged phasing), and the amplitudes of the seasonal cycles (maximum minus minimum).

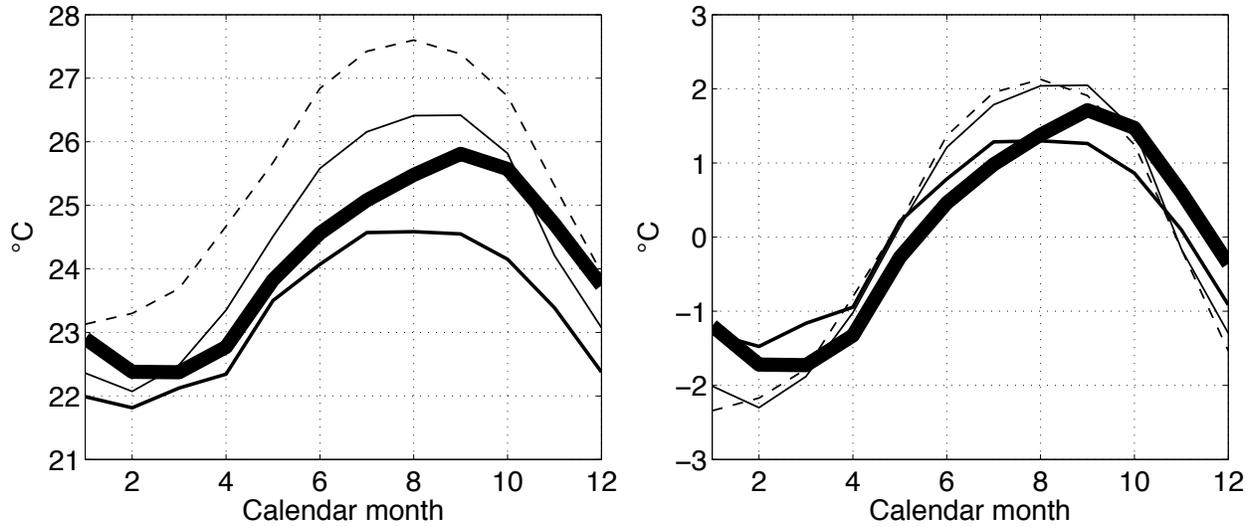

**Figure 8.** Left: Climatological mean seasonal cycles of monthly surface air temperature (°C) at the Hawaiian islands of Kauai (thinnest solid), Oahu (dashed), and Big Island/Island of Hawaii (medium thickness), and at the WHOTS mooring (thickest solid). Correlation coefficients are 0.98, 0.99, and 0.96 for the WHOTS mooring versus Kauai, Oahu, and Big Island/Island of Hawaii, respectively, with the islands leading the mooring by one month in every case. Right: As in left, but with the annual mean removed from each time series.